\documentclass[letterpaper,titlepage,11pt]{article}

\usepackage{amssymb,amsmath,amsfonts}
\usepackage{graphicx}
\usepackage{epsfig}
\usepackage{ccaption}
\usepackage{wrapfig}
\usepackage[small,bf,up]{caption}

\usepackage[
      colorlinks=true,
      linkcolor=black,
      urlcolor=black,
      filecolor=black,
      citecolor=black,
      pdfstartview=FitV,
      pdftitle={},
        pdfauthor={Kentaro Tanabe},
        pdfsubject={},
        pdfkeywords={},
        pdfpagemode=None,
        bookmarksopen=true
      ]{hyperref}

\def\clock{{\count0=\time
           \divide\count0 60
           \ifnum\count0<10 0\fi\the\count0
           \multiply\count0 -60 \advance\count0 \time
           :\ifnum\count0<10 0\fi \the\count0
         }}
\newcommand{\timestamp}{{\small\vbox{\hbox{\tt\jobname.tex}
\hbox{\the\day/\the\month/\the\year, \clock}}}}

\newcommand{\sR}{\mathsf{R}}

\newcommand{\hk}{\hat{\kappa}}

\newcommand{\pz}{\partial_{z}}
\newcommand{\pv}{\partial_{v}}

\numberwithin{equation}{section}

\begin{document}

\begin{titlepage}
\rightline{KEK-TH-1876} 
\leftline{}
\vskip 2cm
\centerline{\LARGE \bf Instability of de Sitter Reissner-Nordstrom black hole }
\medskip
\centerline{\LARGE \bf in the $1/D$ expansion} 
\vskip 1.6cm
\centerline{\bf Kentaro Tanabe$^{a}$}
\vskip 0.5cm
\centerline{\sl $^{a}$Theory Center, Institute of Particles and Nuclear Studies, KEK,}
\centerline{\sl  Tsukuba, Ibaraki, 305-0801, Japan}
\smallskip
\vskip 0.5cm
\centerline{\small\tt ktanabe@post.kek.jp}

\vskip 1.6cm
\centerline{\bf Abstract} \vskip 0.2cm \noindent
We study the large $D$ effective theory for $D$ dimensional charged (Anti) de Sitter black holes. Then we show that the de Sitter Reissner-Nordstrom black hole becomes unstable against gravitational perturbations at larger charge than a critical charge in a higher dimension. Furthermore we find that there is a non-trivial zero-mode static perturbation at the critical charge.  
The existence of static perturbations suggests the appearance of non-spherical symmetric solution branches of static charged de Sitter black holes. 
This expectation is confirmed by constructing the non-spherical symmetric static solutions of large $D$ effective equations. 

\end{titlepage}
\pagestyle{empty}
\small
\normalsize
\newpage
\pagestyle{plain}
\setcounter{page}{1}

\section{Introduction}

The effect of a cosmological constant on black hole properties is quite non-trivial. As one example Konoplya and Zhidenko \cite{Konoplya:2008au} found 
that the de Sitter Reissner-Nordstrom black hole becomes unstable at enough large charge in $D>6$ dimensions (see also \cite{Cardoso:2010rz,Konoplya:2013sba}). This instability of the de Sitter Reissner-Nordstrom black hole is a remarkable property of black hole solution with 
a cosmological constant (for stability of other static black holes, see \cite{Konoplya:2011qq,Ishibashi:2011ws}). Furthermore the paper \cite{Konoplya:2008au}
found that there is a nontrivial zero-mode perturbation at the edge of the instability, and they constructed static deformed solutions of the zero-mode perturbation. This perturbative deformed static solution suggests the existence of a deformed static charged de Sitter black 
hole solution in contrast to the uniqueness of
the static charged asymptotically flat black hole \cite{Gibbons:2002ju}
\footnote{
The instability and static deformed solution of de Sitter Reissner-Nordstrom black hole were also observed independently in the Freund-Rubin solution,
which is the Nariai limit of de Sitter Reissner-Nordstrom black hole
\cite{Kinoshita:2007uk, Kinoshita:2009hh}. 
}.   

The purpose of this paper is to study some properties of charged (Anti) de Sitter black holes such as quasinormal modes and non-linear deformations of 
static solutions by using the large $D$ expansion method \cite{Emparan:2013moa}. Especially we derive large $D$ effective equations for charged (Anti) de Sitter black 
holes by similar way with \cite{Emparan:2015gva,Tanabe:2015hda}\footnote{
The extension of the large $D$ effective theory to asymptotically flat charged black holes has been considered recently in \cite{Bhattacharyya:2015fdk}. 
} in section \ref{2}. These effective equations describe $O(1/D)$ non-linear dynamical deformations of charged black holes. 
Then we obtain an analytic formula for quasinormal mode frequencies of (Anti) de Sitter Reissner-Nordstrom black hole by perturbation analysis 
of the effective equations in section \ref{3}. From this analytic formula the instability modes and their thresholds of de Sitter Reissner-Nordstrom black hole are identified. As found in \cite{Konoplya:2008au} the threshold modes are
static zero-mode perturbations, which suggest the existence of deformed non-spherical symmetric charged de Sitter black hole solutions in higher dimensions. 
In this paper the existence of deformed static charged black hole solutions will be shown explicitly by constructing 
the solutions of large $D$ effective equations in non-linear way in section \ref{4}. These results show that the large $D$ expansion
method is a powerful tool to investigate black hole properties also for charged and non-asymptotically flat solutions.

\section{Large $D$ effective equations}
\label{2}

We derive effective equations for charged (Anti) de Sitter black holes by $1/D$ expansion method. We solve the Einstein-Maxwell equation with a cosmological constant and the Maxwell equation
%
\begin{eqnarray}
R_{\mu\nu}-\frac{1}{2}Rg_{\mu\nu}+\Lambda g_{\mu\nu} = \frac{1}{2}\left( 
F_{\mu\rho}F^{\rho}{}_{\nu}-\frac{1}{4}F_{\rho\sigma}F^{\rho\sigma}g_{\mu\nu}
\right),~~\nabla^{\mu}F_{\mu\nu}=0,
\label{EMeq}
\end{eqnarray}
%
where $F_{\mu\nu}=\partial_{\mu}A_{\nu}-\partial_{\nu}A_{\mu}$. 
The metric ansatz for $D=n+3$ dimensional charged black holes in ingoing Eddington-Finkelstein coordinates is
%
\begin{eqnarray}
ds^{2}=-A dv^{2} +2(u_{v}dv+u_{z}dz)dr -2C_{z}dvdz  + r^{2}G dz^{2} +r^{2}H^{2}d\Omega^{2}_{n}.
\end{eqnarray}
%
The gauge field ansatz is
%
\begin{eqnarray}
A_{\mu}dx^{\mu} = A_{v}dv + A_{z}dz. 
\end{eqnarray}
%
Note that we do not consider $A_{r}dr$ in the gauge field because it does not contribute to effective equations at the leading order
in $1/D$ expansion. 
We introduce a curvature scale, $L$, of a cosmological constant given by
\footnote{
Here we set a cosmological constant to be positive. The following analysis can be applied also to Anti de Sitter black holes 
just by replacing $L\rightarrow i L$. 
}
%
\begin{eqnarray}
\Lambda=\frac{(n+1)(n+2)}{2L^{2}}. 
\end{eqnarray}
%
In this coordinate de Sitter Reissner-Nordstrom black hole is given by
%
\begin{eqnarray}
ds^{2} = -\left( 1 -\frac{r^{2}}{L^{2}}-\left(\frac{r_{+}}{r}\right)^{n}+\tilde{Q}^{2}\left(\frac{r_{+}}{r}\right)^{2n} \right) dv^{2} +2dvdr +r^{2}(dz^{2}+\sin^{2}{z}~d\Omega^{2}_{n}),
\end{eqnarray}
%
and
%
\begin{eqnarray}
A_{\mu}dx^{\mu} = \sqrt{\frac{2(n+1)}{n}}\tilde{Q}\left(\frac{r_{+}}{r}\right)^{n}dv.
\end{eqnarray}
%
$r_{+}$ is a horizon radius parameter, and $\tilde{Q}$ is a charge parameter of the solution. We construct large $D$ effective theory for dynamical deformations 
of this exact solution. To obtain the effective theory, we assume that the radial gradient becomes dominant at large $D$ as $\partial_{v}=O(1)$, $\partial_{z}=O(1)$ and $\partial_{r}=O(n)$. Note that we use $n$ as a 
large parameter instead of $D$ in the following. It is useful to introduce new radial coordinate $\sR$ defined by
%
\begin{eqnarray}
\sR=\left( \frac{r}{r_{0}} \right)^{n},
\end{eqnarray}
%
where $r_{0}$ is a constant parameter of the solution. We set to $r_{0}=1$. Then  $\partial_{r}=O(n)$ corresponds to $\partial_{\sR}=O(1)$. The large $n$ scalings of metric and gauge field functions are
%
\begin{gather}
A=O(1),~u_{v}=O(1),~u_{z}=O(n^{-1}),~C_{z}=O(n^{-1}),~\notag \\
A_{v}=O(1),~~A_{z}=O(n^{-1}). 
\end{gather}
%
The metric functions, $G$ and $H$, are assumed to be
%
\begin{eqnarray}
G=1 +O(n^{-1}),~~H=H(z). 
\end{eqnarray}
%
$u_{z}$ is a shift vector on $r=\text{const.}$ surface, and we choose $u_{z}=u_{z}(v,z)$ as a gauge choice. Other metric functions and gauge 
field functions depend on $(v,r,z)$. The Reissner-Nordstrom black hole satisfies these large $n$ scaling rules. We assume that the deviation from the Reissner-Nordstrom black hole starts at $O(n^{-1})$ in the metric. For example, the Reissner-Nordstrom black hole has $C_{z}=0$, but our solutions considered here have $C_{z}=O(n^{-1})$. 

The leading orders of the field equations (\ref{EMeq}) contain only $\sR$-derivative, and we can solve them easily. As a result we obtain 
following leading order solutions of the field equations in $1/n$ expansion
%
\begin{gather}
A=A_{0}^{2}\left( 1-\frac{m}{\sR}+\frac{q^{2}}{\sR^{2}} \right),~~A_{v} =\frac{\sqrt{2}q}{\sR},~~
u_{v}=\frac{A_{0}L H(z)}{\sqrt{L^{2}(1-H'(z)^{2})-H(z)^{2}}},
\label{LOsol1}
\end{gather}
%
and
%
\begin{eqnarray}
u_{z}=\frac{u^{(0)}_{z}(z)}{n},~~
C_{z}=\frac{1}{n}\Biggl[\frac{p_{z}}{\sR} -\frac{p_{z} q^{2}}{m\sR^{2}}\Biggr],~~
A_{z}=\frac{1}{n}\frac{\sqrt{2}p_{z}q}{m \sR},~~G=1+O(n^{-2}).
\label{LOsol2}
\end{eqnarray}
%
Here we omit terms of $O(1/n)$ for simplicity. 
$m(v,z)$, $q(v,z)$ and $p_{z}(v,z)$ are integration functions of leading order solutions. 
$m(v,z)$ is a mass density, and $q(v,z)$ is a charge density of the solution. $p_{z}(v,z)$ is a momentum density along $z$ direction. 
$u_{z}^{(0)}(z)$ is an arbitrary function of $z$. $A_{0}$
is a constant describing the redshift factor of the background geometry. Although $A_{0}$ can be a function of $z$ in general, 
we set $A_{0}$ to be constant
\footnote{
In \cite{Emparan:2015hwa} some properties of $O(1)$ deformed static solutions with $A_{0}(z)$ has been studied. 
Since we consider small deformation with $O(1/n)$ amplitude in this paper, we can take $A_{0}$ to be constant. At next-to-leading order we have $z$-dependent redshift factor. 
}. We can see that $g_{vv}$ vanishes at $\sR=\sR_{\pm}$ given by
%
\begin{eqnarray}
\sR_{\pm} = \frac{m\pm\sqrt{m^{2}-4q^{2}}}{2} 
\end{eqnarray}
%
in the leading order solution. These surfaces can be regarded as an outer and inner horizon by analogy with the Reissner-Nordstrom black hole. The extremal condition
is $m(v,z)=2q(v,z)$, and we assume $m(v,z)>2q(v,z)\geq 0$ in this paper. 
As boundary conditions we imposed the regularity condition of metric functions at $\sR=\sR_{+}$ and the asymptotic condition at $\sR\gg 1$ 
%
\begin{eqnarray}
A_{v}=O(\sR^{-1}),~~A_{z}=O(\sR^{-1}),~~C_{z}=O(\sR^{-1}). 
\end{eqnarray}
%
The momentum constraint of the Einstein-Maxwell equation on $r=\text{const.}$
surface at the leading order gives
%
\begin{eqnarray}
\frac{d}{d z} \frac{ H(z)}{\sqrt{L^{2}(1-H'(z)^{2})-H(z)^{2}}}=0.
\label{Hcond}
\end{eqnarray}
%
Using this constancy condition we define a constant $\hat{\kappa}$ by
%
\begin{eqnarray}
\hat{\kappa} = \frac{A_{0}\sqrt{L^{2}(1-H'(z)^{2})-H(z)^{2}}}{2LH(z)}.
\end{eqnarray}
%
The constant $\hat{\kappa}$ has a relation with the surface gravity of static solutions as we will see below. 

Note that $O(n^{0})$ parts of the leading order solutions in eqs. (\ref{LOsol1}) and (\ref{LOsol2}) have same form with one of the Reissner-Nordstrom black hole except for $(v,z)$ dependences in $m$ and $q$. This reminds us of the Birkhoff theorem for spherical symmetric solutions in the Einstein-Maxwell system. Actually the radial dynamics becomes dominant at large $n$, and, as a result, the leading order solutions at large $n$ have same structure in the radial dynamics with spherical symmetric spacetimes. This feature can be observed also in the rotating black holes \cite{Emparan:2013xia}, and the paper \cite{Bhattacharyya:2015fdk} used this feature to construct the effective theory of general charged black holes at large $D$.   

At the next-to-leading order of the field equations (\ref{EMeq}), we obtain conditions for $m(v,z)$, $q(v,z)$ and $p_{z}(v,z)$ as the effective 
equations for charged (Anti) de Sitter black hole. The effective equations are
%
\begin{eqnarray}
\pv q - \frac{A_{0}^{2}H'(z)}{2\hat{\kappa}H(z)}\pz q+\frac{H'(z)}{H(z)}\frac{p_{z}q}{m}=0,
\label{eq1}
\end{eqnarray}
%
%
\begin{eqnarray}
\pv m - \frac{A_{0}^{2}H'(z)}{2\hat{\kappa}H(z)}\pz m+\frac{H'(z)}{H(z)}p_{z}=0,
\label{eq2}
\end{eqnarray}
%
and
%
\begin{eqnarray}
&&
\pv p_{z} -\frac{A_{0}^{2}H'(z)}{2\hat{\kappa}H(z)}\frac{\sqrt{m^{2}-4q^{2}}}{m} \pz p_{z} 
+\Biggl[
A_{0}^{2} - \frac{A_{0}^{2}H'(z)}{2\hat{\kappa}H(z)}\frac{p_{z}(m-\sqrt{m^{2}-4q^{2}})}{m^{2}}
\Biggr]\pz m \notag \\
&&~~~
-\Biggl[
\frac{A_{0}^{2}(1-H'(z)^{2})}{2\hat{\kappa}H(z)^{2}}-\frac{A_{0}^{2}H'(z)^{2}}{2\hat{\kappa}H(z)^{2}}
\frac{\sqrt{m^{2}-4q^{2}}}{m} -\frac{H'(z)}{H(z)}\frac{p_{z}}{m} 
\Biggr]p_{z}=0.
\label{eq3}
\end{eqnarray}
%
The contribution of a cosmological constant enters only through $\hat{\kappa}$ and $A_{0}$. 
We can find a static solution of effective equations by assuming $m(v,z)=m(z)$, $q(v,z)=q(z)$ and $p_{z}(v,z)=p_{z}(z)$. 
From eqs. (\ref{eq1}) and (\ref{eq2}), $p_{z}(z)$ and $q(z)$ are found to be
%
\begin{eqnarray}
p_{z}(z)= \frac{A_{0}^{2}m'(z)}{2\hat{\kappa}},~~q(z)=\tilde{Q} m(z).
\label{ssol}
\end{eqnarray}
%
$\tilde{Q}$ is an integration constant, and $q$ should be proportional to $m$. 
Eq. (\ref{eq3}) gives a simple equation for $m(z)=e^{P(z)}$ as
%
\begin{eqnarray}
P''(z) -\Biggl[
\frac{H'(z)}{H(z)} -\frac{H(z)}{L^{2}\sqrt{1-4\tilde{Q}^{2}}}
\Biggr]P'(z)=0.
\label{Peq}
\end{eqnarray}
%
The static solution has the Killing vector $\pv$. The associated surface gravity, $\kappa$, 
of this static solution at the outer horizon $\sR=\sR_{+}$ is calculated as
%
\begin{eqnarray}
\kappa = n\hat{\kappa}\frac{\sqrt{1-4\tilde{Q}^{2}}-(1-4\tilde{Q}^{2})}{2\tilde{Q}^{2}}\bigl[1+O(n^{-1})\bigr].
\end{eqnarray}
%
Hence the static solution has a constant surface gravity as expected. To solve eq. (\ref{Peq}), we need explicit form of $H(z)$. The function $H(z)$ can be specified if we consider an embedding of the leading order solution into a background geometry. As a condition
for the embedding, $H(z)$ should satisfy the condition  (\ref{Hcond}).

Note that we have considered static solutions of effective equations. This static solution of effective equations describe also the static spacetime. As seen in eq. (\ref{ssol}) the momentum $p_{z}$ is given by total derivative, so there is no global momentum along $z$ direction. This means that the spacetime is also static.   

\section{Instability of de Sitter Reissner-Nordstrom black hole}
\label{3}

We apply effective equations to the stability analysis of (Anti) de Sitter Reissner-Nordstrom black hole. 
(Anti) de Sitter Reissner-Nordstrom black hole is realized as a static solution of effective equations in the embedding
into de Sitter spacetime in the spherical coordinate. 
The embedding in the spherical coordinate is given by
%
\begin{eqnarray}
H(z)=\sin{z}. \label{Hcond2}
\end{eqnarray}
%
This embedding satisfies the condition for $H(z)$ in eq. (\ref{Hcond}). 
The redshift factor $A_{0}$ is set to be
%
\begin{eqnarray}
A_{0}=\sqrt{1-\frac{1}{L^{2}}},
\end{eqnarray}
%
and we consider the small de Sitter black hole $L>1$. In this embedding the constant $\hk$ becomes
%
\begin{eqnarray}
\hk=\frac{L^{2}-1}{2L^{2}}.
\label{hkeq}
\end{eqnarray}
%
Then de Sitter Reissner-Nordstrom black hole is given by a static solution
%
\begin{eqnarray}
m(v,z)=1+Q^{2},~~q(v,z)=Q,~~p_{z}(v,z)=0.
\label{RNsol}
\end{eqnarray}
%
$Q$ is a charge parameter of the de Sitter Reissner-Nordstrom black hole. In this solution we set the horizon radius to be unity as $\sR_{+}=1$. In this unit
the extremal limit is $Q=1$. The relation of $Q$ and $\tilde{Q}$ in eq. (\ref{ssol}) are
%
\begin{eqnarray}
\tilde{Q} = \frac{Q}{1+Q^{2}}.
\end{eqnarray}
%
By perturbing effective equations (\ref{eq1}), (\ref{eq2}) and (\ref{eq3}) around the static solution (\ref{RNsol}) as
%
\begin{eqnarray}
m(v,z)=(1+Q^{2})(1+\epsilon e^{-i\omega v}F_{m}(z)),~~q(v,z)=Q(1+\epsilon e^{-i\omega v}F_{q}(z)),
\end{eqnarray}
%
and
%
\begin{eqnarray}
p_{z}(v,z)=\epsilon e^{-i\omega v}F_{z}(z),
\end{eqnarray}
%
we obtain quasinormal modes of de Sitter Reissner-Nordstrom black hole given by
%
\begin{eqnarray}
\omega_{\pm} = -\frac{i(\ell-1)}{1+Q^{2}} \pm \frac{\sqrt{\ell(L^{2}(1+Q^{4})-(1+Q^{2})^{2})-L^{2}(1+\ell^{2}Q^{4})}}{L(1+Q^{2})},
\label{grQNMs}
\end{eqnarray}
%
for the gravitational perturbation ($F_{m}(z)=F_{q}(z)$), and
%
\begin{eqnarray}
\omega = -i\ell,
\label{gaQNMs}
\end{eqnarray}
%
for the gauge field perturbation ($F_{m}(z)\neq F_{q}(z)$). $\ell=0,1,2,...$ is the quantum number on $S^{n+1}$. 
As boundary condition we demanded $F_{m}(z)\propto (\cos{z})^{\ell}$ at $z=\pi/2$
\footnote{
This boundary condition comes from the behavior of the spherical harmonics at large $D$ \cite{Suzuki:2015iha}. 
}. 
The modes with $\ell>1$ ($\ell>0$) describe dynamical perturbation of gravitational perturbations (gauge field perturbations). So we consider the gravitational perturbations with $\ell>1$ and gauge field perturbations with $\ell>0$.  
At the zero cosmological constant limit $L\rightarrow \infty$, these results reproduce the
quasinormal modes of Reissner-Nordstrom black hole obtained in \cite{Bhattacharyya:2015fdk}. At $Q=0$, $\omega_{\pm}$ reduces to quasinormal modes of
scalar type perturbation of (Anti) de Sitter Schwarzschild black hole \cite{Emparan:2015rva}. 
At the both limits, the perturbations are stable. On the other hand 
we can see that, as a striking effect of a positive cosmological constant, the gravitational perturbation becomes unstable at $Q>Q_{\ell}$ where
%
\begin{eqnarray}
Q_{\ell}=\sqrt{\frac{L^{2}(\ell-1)-1}{L^{2}(\ell-1)+1}}.
\label{Qell}
\end{eqnarray}
%
%
\begin{figure}[t]
 \begin{center}
  \includegraphics[width=65mm,angle=0]{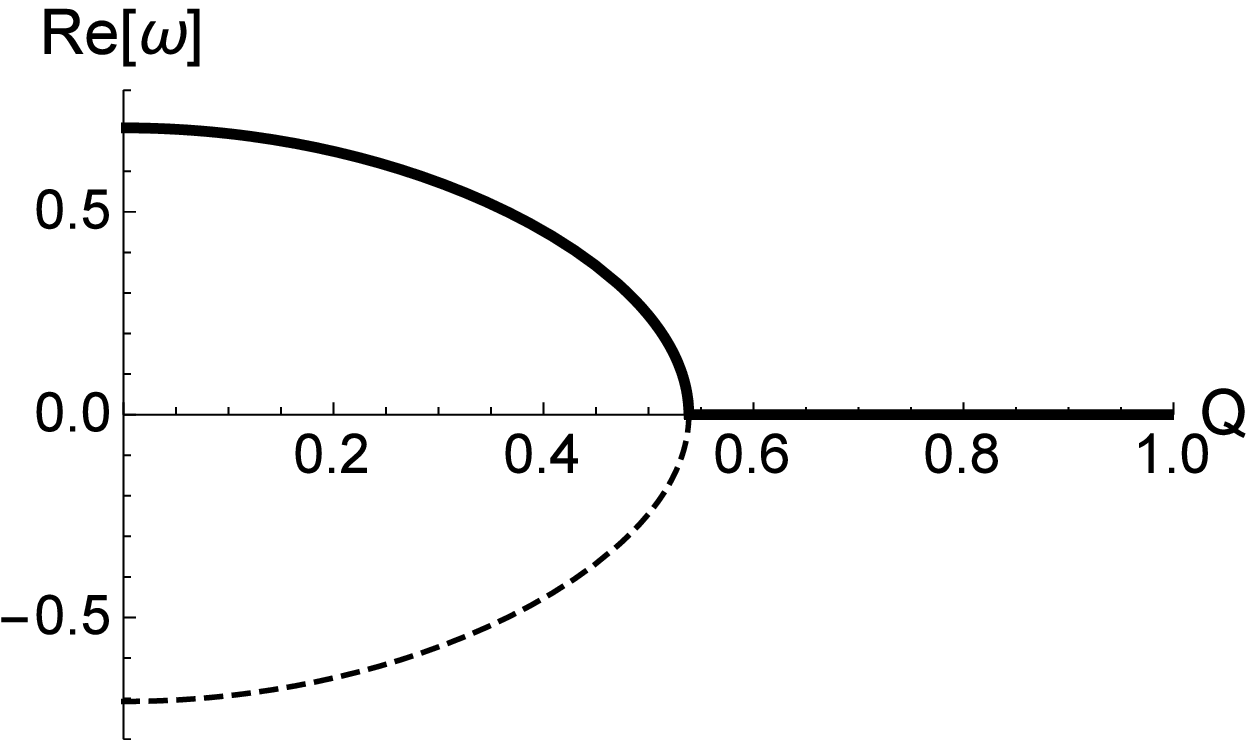}
  \hspace{5mm}
  \includegraphics[width=65mm,angle=0]{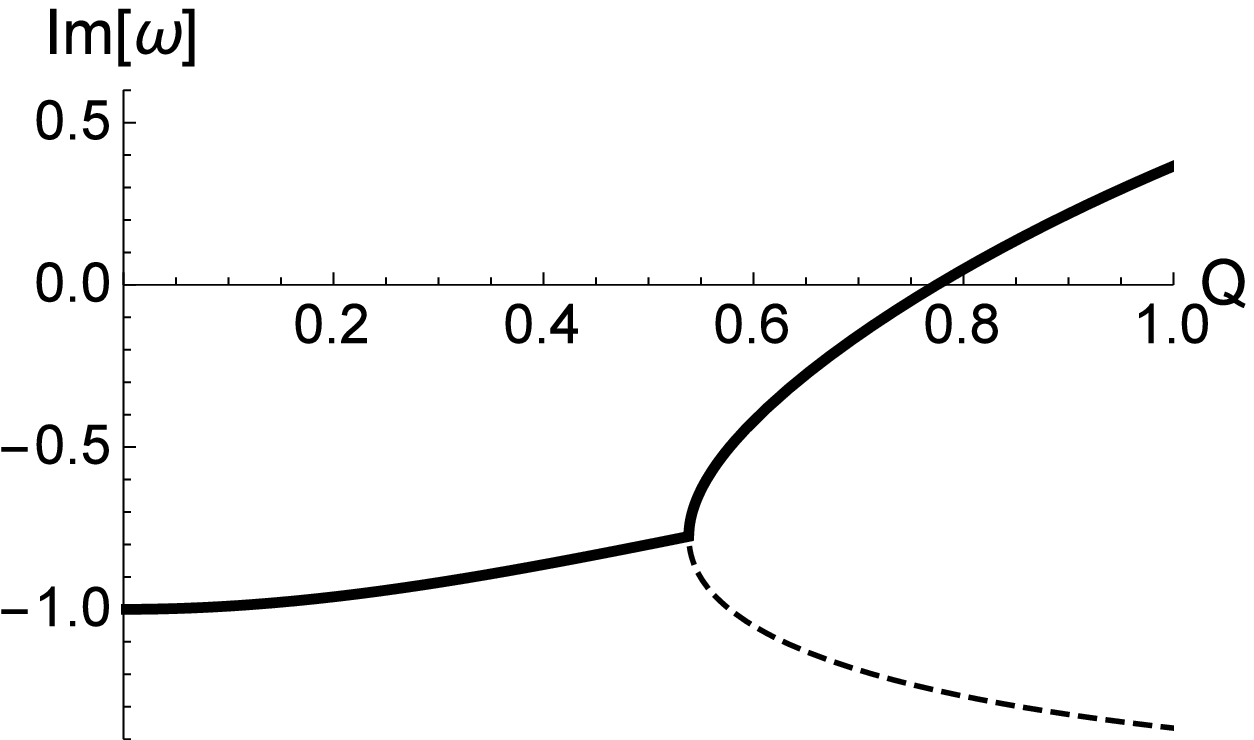}
 \end{center}
 \vspace{-5mm}
 \caption{Plots of the quasinormal mode frequency, $\omega_{+}$ (thick line) and $\omega_{-}$ (dashed line), of the gravitational perturbation of de Sitter Reissner-Nordstrom black hole with $\ell=2$ and $L/r_{0}=2$. 
The left panel shows the real part, and right panel shows the imaginary part. At $Q>Q_{\ell=2}\sim0.775$ the perturbation becomes unstable.}
 \label{fig1}
\end{figure}
%
We can see $Q_{\ell}\geq Q_{\ell=2}$. Thus the most unstable mode is $\ell=2$. 
Furthermore $\omega_{\pm}$ is a purely imaginary mode at $Q>Q_{c}$ where
%
\begin{eqnarray}
Q_{\ell}>Q_{c} \equiv \Biggl[ \frac{(\ell-1)L\sqrt{\ell(L^{2}-1)}-\ell}{\ell((\ell-1)L^{2}+1)} \Biggr]^{1/2}.
\end{eqnarray}
%
So the instability mode is always a purely imaginary mode, and there is a non-trivial zero-mode static perturbation at $Q=Q_{\ell}$. Then we expect that there is a non-spherical symmetric static solution branch at $Q=Q_{\ell}$.
In Figure.\ref{fig1} we show plots of gravitational quasinormal modes $\omega_{\pm}$ of de Sitter Reissner-Nordstrom black hole. 
%
\begin{figure}[t]
 \begin{center}
  \includegraphics[width=80mm,angle=0]{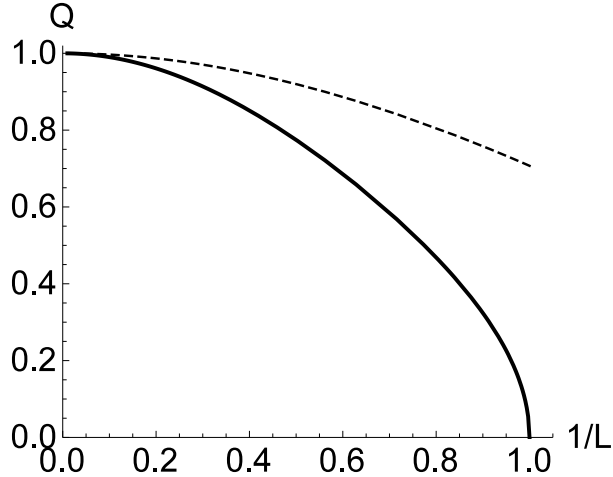}
 \end{center}
 \vspace{-5mm}
 \caption{  
The unstable region is shown. The thick line is $Q_{\ell}$ of $\ell=2$, and the dashed line is of $\ell=4$. The region above each line (larger charge) is unstable region for each mode perturbation.}
 \label{fig2}
\end{figure}
%
In Figure.\ref{fig2} we show the unstable region of the de Sitter Reissner-Nordstrom black hole in $(L,Q)$ plane. The thick line describes $Q_{\ell}$ of $\ell=2$, and the dashed line is of $\ell=4$. The region above each line is unstable region for each mode perturbation. The unstable region seems to extend to $Q=0$ at $L=1$. However, since our solution breaks down at the extremal limit $L=1$ due to $\hat{\kappa}=0$ (see eq. (\ref{hkeq})), our result does not immediately imply that the de Sitter Schwarzschild black hole is unstable at the Nariai limit $L=1$. The stability at the Narial limit was studied in \cite{Kinoshita:2007uk,Kinoshita:2009hh}.

In the paper \cite{Konoplya:2013sba} an analytic formula for the critical charge in arbitrary dimensions was given. However the formula is not same with our formula (\ref{Qell}) even at large $D$. The analytic formula in \cite{Konoplya:2013sba} was obtained by extrapolating their numerical results in $D=6\sim 11$. However there are nonperturbative contributions in $1/D$ expansions with $O(e^{-D})$ \cite{Emparan:2015rva}, and such nonperturbative corrections cannot be ignored in not so much higher dimensions such as $D=6$. So, if one uses numerical results in such dimensions to predict the $1/D$ expanded formula, the correct formula in $1/D$ expansion would not be obtained. 

\paragraph{AdS Reissner-Nordstrom black hole} 

Replacing $L$ by $i L$, we can obtain effective equations for charged Anti de Sitter black holes. Especially we can obtain quasinormal modes of the Anti de Sitter Reissner-Nordstrom black hole from eqs (\ref{grQNMs}) and (\ref{gaQNMs}) by the replacing of $L\rightarrow iL$. These quasinormal modes does not show any instabilities as we can see in eq. (\ref{Qell}). In fact
we have $Q_{\ell}>1$ for Anti de Sitter case. So the black hole is stable unless we consider overcharged solutions. This is consistent with numerical results \cite{Konoplya:2008rq}.  

\section{Deformed solution branches}
\label{4}

The perturbation analysis suggests an appearance of a non-spherical symmetric static charged de Sitter black hole solution branch at $Q=Q_{\ell}$. This expectation can be confirmed directly 
by solving effective equations for static solutions. 
The equation for the static solution is given in eq. (\ref{Peq}). In the spherical coordinate embedding (\ref{Hcond2}), the solution is obtained as
%
\begin{eqnarray}
P(z) = m_{0} + b (\cos{z})^{k(Q)},
\end{eqnarray}
%
where
%
\begin{eqnarray}
k(Q) = 1+\frac{1+Q^{2}}{L^{2}(1-Q^{2})}.
\end{eqnarray}
%
$b$ is an integration constant, and it describes an $O(1/n)$ deformation amplitude from spherical symmetry. $m_{0}$ is an $O(1/n)$ redefinition of the horizon 
radius $r_{0}$. For general $Q$, $k(Q)$ takes non-integer value, and the solution is not analytic at $z=\pi/2$. However 
we obtain $k(Q)=\ell$ at $Q=Q_{\ell}$ with $\ell=0,1,2,...$. Then the solution becomes regular at $z=\pi/2$ for $Q=Q_{\ell}$. $\ell=0$ and $\ell=1$ are just gauge modes in the spherical coordinate embedding, 
so they do not describe physical deformations. For $\ell>1$, the solution gives a static solution branch of non-spherical symmetric charged de Sitter black hole.
Note that this deformation is similar with deformations of the Myers-Perry black hole observed in \cite{Suzuki:2015iha}, that is, the bumpy black holes. 
So it would be interesting to study their relations to clarify the nature of instabilities of de Sitter Reissner-Nordstrom black hole. 


\section{Summary}

We solved the Einstein-Maxwell equation with a cosmological constant and Maxwell equations for charged black holes by using large $D$ expansion method. As a result we obtained effective equations for  charged (Anti) de Sitter black holes. The effective equations describe the non-linear dynamical deformations of de Sitter Reissner-Nordstrom black holes. The perturbation analysis gives
the analytic formula for quasinormal modes of (Anti) de Sitter Reissner-Nordstrom black hole. From this formula we found that the gravitational perturbation of de Sitter Reissner-Nordstrom black hole becomes unstable, and the quasinormal mode is purely imaginary in unstable region. At the onset of instabilities there is a non-trivial zero-mode static perturbation. We obtained corresponding non-linear deformed solution at the leading order in $1/D$ expansion. The non-linear solutions give solution branches of non-spherical symmetric static charged de Sitter black holes in higher dimensions.  

There are various interesting developments and extensions of this work such as construction of phase diagram of deformed solutions obtained in this paper, inclusion of $1/D$ corrections to compare numerical results more accurately or rotation to study the instabilities of charged rotating de Sitter black hole. The charged rotating black hole has not been found analytically even in asymptotically flat case, so such studies would give fruitful results. The $1/D$ corrections
introduce the dimensional dependences to the results, so we may be able to observe a critical dimension, where the instability of de Sitter Reissner-Nordstrom black hole disappear as done in analysis for the critical dimensions of non-uniform black string \cite{Suzuki:2015axa}. 
It would be also quite interesting to solve effective equations for charged de Sitter black holes numerically. The effective equations can 
describe non-linear dynamical evolution of the instabilities of de Sitter Reissner-Nordstrom black hole. The numerical solutions of effective equations will 
give deeper understanding on the instability and its endpoint.

\section*{Acknowledgments}
The author is very grateful to Roberto Emparan for useful discussions and valuable comments on the draft. 
This work was supported by JSPS Grant-in-Aid for Scientific Research No.26-3387.




\end{document}